\documentclass[11pt]{article}
\usepackage{moriond,epsfig}

\def\be{\begin{equation}}
\def\ee{\end{equation}}
\def\bea{\begin{eqnarray}}
\def\eea{\end{eqnarray}}

\def\ppbar{p\bar{p}}
\def\ttbar{t\bar{t}}
\def\mtt{M_{t\bar{t}}}
\def\afb{A_{FB}}
\def\invfb{fb$^{-1}$}
\def\dy{\Delta y}
\begin{document}
\vspace*{4cm}
\title{MEASUREMENTS OF TOP QUARK PROPERTIES AT THE TEVATRON}

\author{ D. MIETLICKI \\
(On Behalf of the CDF and D0 Collaborations)}

\address{Department of Physics, University of Michigan,\\
450 Church Street, Ann Arbor, MI, USA}

\maketitle\abstracts{
The top quark is the most recently discovered of the standard model quarks, and studies of its properties are important 
tests of the standard model.  Many measurements of top properties have been produced by the CDF and D0 
collaborations, which study top quarks produced in $\ppbar$ collisions at the Fermilab Tevatron with a center-of-mass 
energy $\sqrt{s} =  1.96$~TeV.  We describe recent results from top properties measurements at the Tevatron 
using datasets corresponding to integrated luminosities up to $8.7$~\invfb.}

\section{Introduction}

Since its discovery in 1995 at the Tevatron,~\cite{topdisc} the study of the production and properties 
of the top quark has been one of the most active areas of research in high energy particle physics.  Its 
extremely large mass makes the top unique among quarks, in that it decays via the electroweak interaction 
before hadronization.  Thus, properties such as spin can be measured through their effects on the kinematic distributions 
of the top decay products.  This offers physicists a first chance to study a ``bare'' quark.  With a Yukawa 
coupling near one, the top quark could play a special role in electroweak symmetry breaking, 
and its large mass could potentially lead to enhanced couplings to new physics.  
Precision measurements of top properties are important both as tests of the standard model and as 
potential avenues for the discovery of new physics.

The majority of top quarks analyzed at the Tevatron are produced as $\ttbar$ pairs via the strong interaction, 
although they are also produced singly via the electroweak interaction, a mode that was not observed until 
2009.~\cite{stopdisc}  The standard model predicts that tops decay almost always via $t \rightarrow Wb$, so the 
decay modes of $\ttbar$ pairs are described by the two $W$ boson decays.  Two decay modes are used in the analyses 
described in this document: the ``dilepton'' mode, where both $W$'s decay to a lepton (electron or muon) and a neutrino, 
and the ``lepton plus jets'' mode, where one $W$ decays leptonically and the other decays to a pair of quarks.  A large 
portion of $\ttbar$ pairs decay into the ``all-hadronic'' channel, where both $W$'s decay to quark pairs, but this 
channel faces a very large background from QCD multi-jet production and is difficult to use in top properties 
measurements.

\section{Asymmetry in $\ttbar$ Production}

When top pairs are produced in $\ppbar$ collisions, the standard model predicts a small asymmetry, $\mathcal{O}$(7\%) at 
next-to-leading order (NLO), in the number of top quarks that travel along the proton direction compared to the 
number that travel along the antiproton direction.  Using approximately $5$~\invfb, analyses at CDF and D0  
found asymmetries exceeding the prediction with moderate significance.~\cite{afb5}  The CDF result in the lepton plus 
jets channel also observed an increase in the asymmetry as a function of the invariant mass of the $\ttbar$ 
system ($\mtt$).  The measurement of this asymmetry with 
the full Tevatron dataset is an important update and will be complementary to similar measurements at the Large Hadron 
Collider (LHC), where the effect must be measured in a different manner due to the symmetric $pp$ initial state.

CDF has recently released a new measurement of the asymmetry using $8.7$~\invfb, corresponding to the full Tevatron 
dataset.~\cite{cdfafb}  The asymmetry $\afb$, defined in Equation~\ref{eq:afb}, is measured using the frame-invariant 
variable $\dy$, the difference in rapidities between the top quark and the antitop quark.  After 
removing the background contribution and correcting for acceptances and 
detector resolution effects, CDF measures a parton level asymmetry of $(16.2 \pm 4.7)\%$, compared to the prediction 
of 6.6\% determined using the NLO event generator {\sc powheg}~\cite{powheg} with a flat correction applied to account 
for electroweak contributions to the asymmetry.~\cite{ewcorr}

\begin{equation}
\afb = \frac{N(\dy > 0) - N(\dy < 0)}{N(\dy > 0) + N(\dy < 0)}
\label{eq:afb}
\end{equation}

\begin{figure}
\begin{center}
\psfig{figure=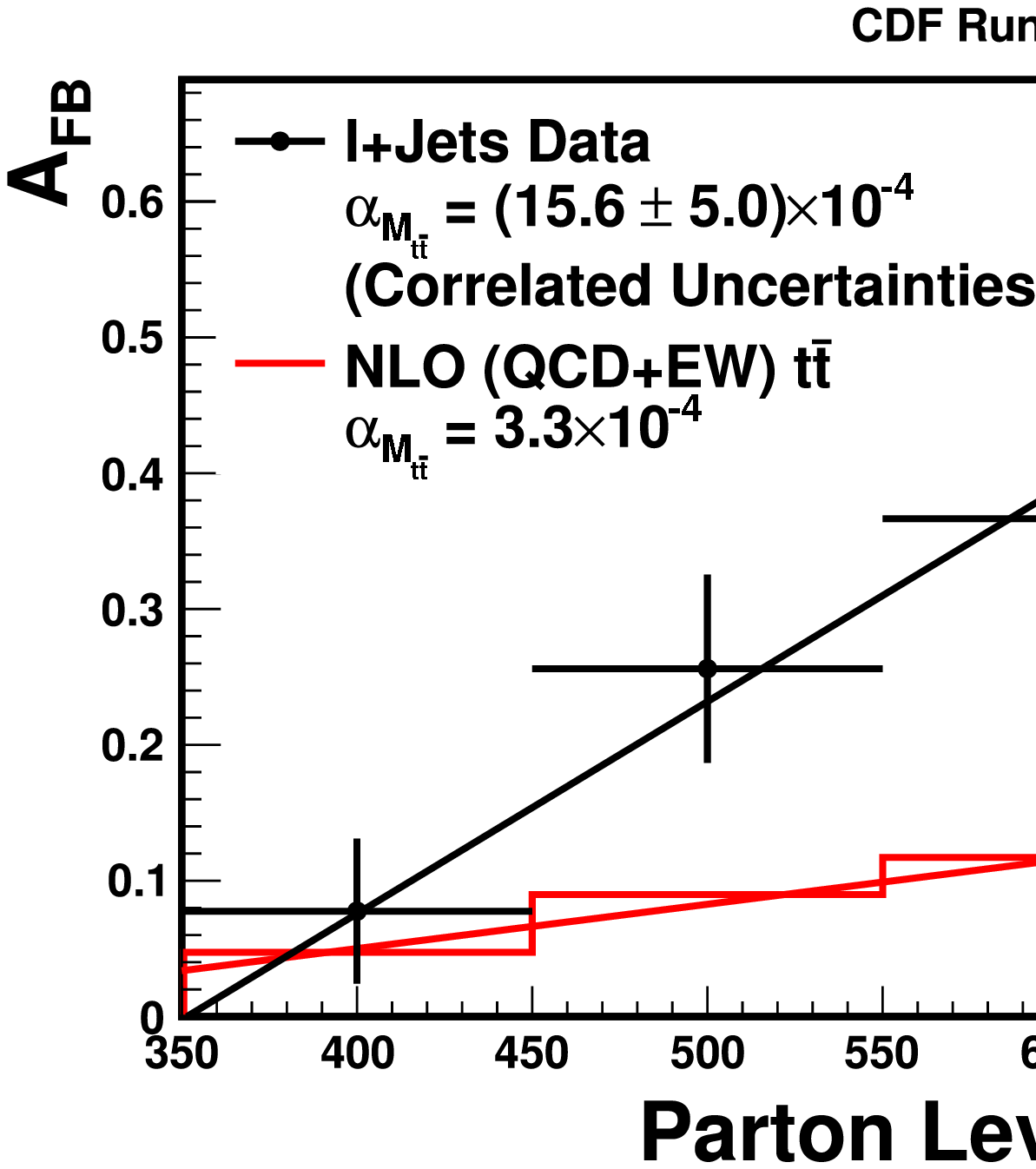,height=2.0in}
\psfig{figure=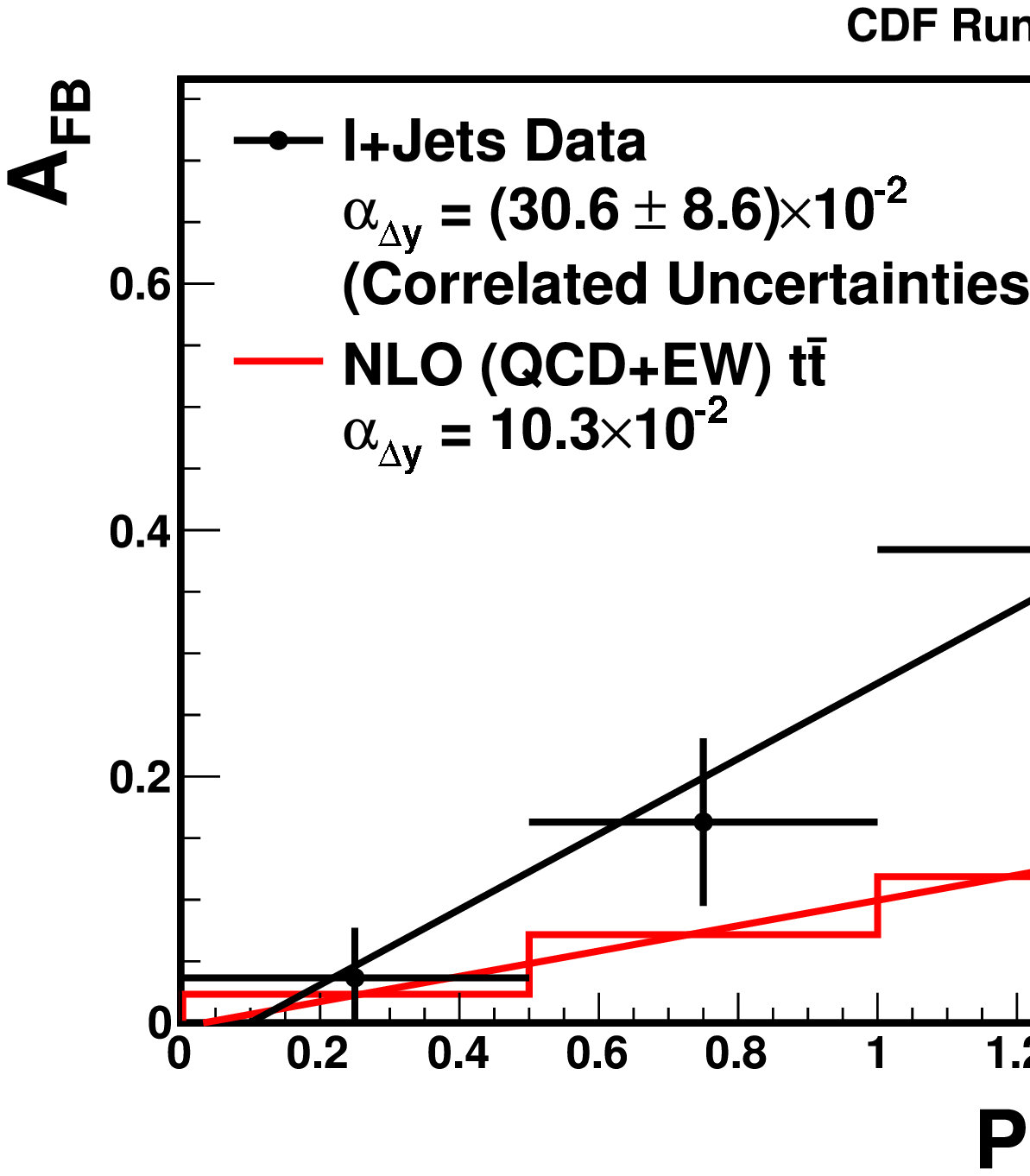,height=2.0in}
\end{center}
\caption{Parton-level $\afb$ as a function of $\mtt$ (left) and $|\dy|$ (right) as measured at CDF.
\label{fig:diffafb}}
\end{figure}

The new CDF result also considers the mass and rapidity dependence of the asymmetry, 
measuring $\afb$ as a function of both $\mtt$ and $|\dy|$, as shown in Figure~\ref{fig:diffafb}.  In both cases, 
the asymmetry is found to be well-fit by a linear ansatz, and the best-fit slopes are measured and compared 
to the prediction.  At the parton level, CDF finds the slope of $\afb$ vs. $\mtt$ to be 
$(15.6 \pm 5.0) \times 10^{-4}$, compared to a prediction of $3.3 \times 10^{-4}$.  The slope of $\afb$ vs. 
$|\dy|$ is measured to be $(30.6 \pm 8.6) \times 10^{-2}$, compared to a prediction of $10.3 \times 10^{-2}$.  The 
significance of the deviation from the prediction is determined at the background-subtracted level, before the final 
corrections for acceptance and resolution effects are applied.  Simulated experiments are performed based on the 
{\sc powheg} prediction with electroweak corrections, and a p-value is determined by finding the fraction of such 
experiments in which the predicted slope is at least as large as that observed in the data.  A p-value of 
$6.46 \times 10^{-3}$ is found for the $\mtt$ dependence of the asymmetry, along with a p-value of 
$8.92 \times 10^{-3}$ for the $|\dy|$ dependence.

\section{$\ttbar$ Spin Correlations}

When top quark pairs are produced in hadronic collisions, the individual top quarks are unpolarized, but the $\ttbar$ 
system has a definite spin state and thus the spins of the two quarks are correlated.  The strength of this correlation, 
which is frame-dependent, is quantified as the fractional difference $\kappa$ between the number of top pairs where the 
quark spins are aligned and the number of pairs where the spins are oppositely aligned.  Because tops decay before 
hadronization, spin information can be measured by considering the angular distributions of the top decay products.  

Both CDF and D0 have performed measurements of the spin correlation, with the spin quantization axis being defined 
to be along the beam direction, where the standard model predicts $\kappa = 0.78$.~\cite{spintheory} CDF uses template 
fitting methods to measure $\kappa$ directly, finding $\kappa =  0.72 \pm 0.69$ in the lepton plus 
jets decay channel~\cite{cdfljspin} and $\kappa =  0.042 \pm 0.563$ in the dilepton channel.~\cite{cdfdilspin}  
The CDF results are consistent with each other and with the standard model within large uncertainties.

D0 utilizes a new matrix element approach that enhances the sensitivity by approximately 30\%.  A matrix element 
method is used to define a discriminant based on the probability that a given event contains the standard model 
spin correlation.  Combining measurements in the dilepton~\cite{d0dilspin} and lepton plus jets~\cite{d0ljspin} 
channels, D0 finds that the fraction of events which contain the standard model spin correlation is 
$f = 0.85 \pm 0.29$.  This result provides the first $3\sigma$ evidence for the existence of the 
spin correlation.  The fraction of events containing the standard model correlation 
is then converted to a measurement of  $\kappa$, giving $\kappa =  0.66 \pm 0.23$.

\section{$W$ Boson Helicity}

In the standard model, top quarks decay nearly always to a $W$ boson and a $b$ quark, and the 
helicity states of the $W$ are constrained 
according the the V-A nature of the $W-t-b$ coupling.  The standard model predicts that the fractions of longitudinal, 
left-handed, and right-handed $W$ bosons, labeled $f_{0}$,$f_{-}$, and $f_{+}$ respectively, in $\ttbar$ events will 
be approximately 0.7, 0.3, and 0.0.

D0 and CDF have both performed measurements of these helicity fractions by considering angular 
distributions of the $W$ decay products - particularly the lepton -  in $\ttbar$ candidate events.  A 
combination of the $W$ helicity results for the two experiments has recently been submitted for 
publication,~\cite{whelcombo} the first such published combination of $W$ helicity measurements.  
With $f_{0} + f_{-} + f_{+} = 1$, the combined CDF and D0 measurements find $f_{0} = 0.722 \pm 0.081$ 
and $f_{+} = -0.033 \pm 0.046$.

\section{Top Branching Ratio \label{sec:branching}}

The standard model prediction that top quarks almost always decay to $Wb$ has also been tested in 
measurements by CDF and D0.  Both collaborations have recently performed analyses to measure the 
ratio $R = \mathcal{B}(t \rightarrow Wb)/\mathcal{B}(t \rightarrow Wq)$, where $\mathcal{B}(t \rightarrow WX)$ 
is the branching ratio for a top to decay to $WX$.  The standard model predicts that $R$ should be 
very close to 1.  If the CKM matrix is assumed to be unitary, then a measurement of $R$ can also 
be converted into a measurement of the CKM matrix element $|V_{tb}|$.

In $\ttbar$ production and decay, the standard model expectation is that each event will contain two $b$ quarks.  
Since jets originating from $b$ quarks can be tagged by a displaced secondary vertex, and the efficiency 
for tagging such jets can be measured, the ratio $R$ can be determined by dividing the sample of $\ttbar$ 
candidate events into sub-samples with 0, 1, or 2 $b$-tagged jets and comparing the relative sizes of 
each sub-sample to the predicted sizes determined from the tagging efficiency.  Using a luminosity of $5.4$~\invfb, a 
recent D0 measurement in both the dilepton and lepton plus jets channels~\cite{d0branching} 
finds $R = 0.90 \pm 0.04$, smaller than the standard model expectation at the level of approximately 
$2\sigma$, and measures $|V_{tb}| = 0.95 \pm 0.02$.  With $7.5$~\invfb, a new CDF result in the 
lepton plus jets channel~\cite{cdfbranching} measures $R = 0.91 \pm 0.09$ and $|V_{tb}| = 0.95 \pm 0.05$, 
again somewhat below the prediction but with a significance that is smaller than the D0 result.

\section{Top Width}

The top quark width is expected to be approximately 1.5 GeV in the standard model, and both CDF and D0 
have performed measurements to test this prediction.  CDF has performed a direct 
measurement in the lepton plus jets decay channel with a luminosity of $4.3$~\invfb, 
using a likelihood fit to to the reconstructed top quark mass distribution based on template 
samples with different input top widths.~\cite{cdfwidth}  This analysis results in a 95\% C.L. limit of 
$\Gamma_{t} < 7.6$~GeV and a 68\% two-sided limit $0.3$~GeV~$ < \Gamma_{t} < 4.4$~GeV.

Using a luminosity of $5.4$~\invfb, D0 has performed a complementary measurement that indirectly 
measures the top width by combining results from other top properties 
measurements.~\cite{d0width}  In particular, as shown in Equation~\ref{eq:width}, the total width of 
the top quark is determined from the ratio of the partial width for the process $t \rightarrow Wb$, 
as determined from the measured cross-section for single top production, to the branching ratio 
for $t \rightarrow Wb$, measured in the analysis described in Section~\ref{sec:branching}.  This method 
requires input from several measurements and from the theoretical predictions, but results in 
increased sensitivity compared to a direct measurement.  D0 measures a width of 
$\Gamma_{t} = 2.00^{+0.47}_{-0.43}$~GeV, and converts this to a 95\% C.L. 
limit on $|V_{tb}|$, finding $0.81 < |V_{tb}| \le 1$.

\begin{equation}
\Gamma_{t} = \frac{\Gamma(t \rightarrow Wb)}{\mathcal{B}(t \rightarrow Wb)}
\label{eq:width}
\end{equation}

\section{Conclusions}

The full dataset collected at the Tevatron is now being used to measure top quark properties at CDF and D0.  Many 
of these measurements, such as the $\afb$ measurement and the measurement of $\ttbar$ spin correlations, are 
complementary to analyses that can be performed at the LHC, where the different center-of-mass energy and initial 
state will provide additional information about the couplings of the top quark.  Many CDF and D0 analyses, such as 
the $W$ helicity measurement in $\ttbar$ decays described here, are now being combined to create Tevatron-wide 
results.  Data-taking has ceased at the Tevatron, but there is still much left to be learned from analysis of the 
top quark samples collected at CDF and D0, and both collaborations continue to pursue precision results.

\section*{Acknowledgments}
I would like to thank the members of the CDF and DO collaborations for their dedication to the study of
top quark properties, especially the top group conveners and the authors of the analyses included here, as 
well as the CDF and D0 funding agencies who have made such research possible.  I also want to thank the organizers 
of the 2012 Rencontres de Moriond and all contributors.

\end{document}